\documentclass[conference]{ieeeconf}  

\IEEEoverridecommandlockouts                              

\usepackage{graphicx}
\usepackage{epstopdf}
\usepackage{amsfonts}
\usepackage{amsmath,amssymb}
\usepackage{color}
\usepackage{fancyheadings}
\usepackage{tikz}
\usetikzlibrary{automata}
\usepackage{color}
\usepackage{verbatim}
\usepackage{algorithm}
\usepackage{subcaption}
\usepackage{booktabs}
\usepackage{tabularx}
\usepackage{rotating}
\usepackage{caption,subcaption}
\usepackage[switch]{lineno}
\usepackage{multirow}
\usepackage[autostyle]{csquotes}
\usepackage{hyperref} 
\hypersetup{
    urlcolor=blue    
}

\usepackage{setspace}
\begin{document}
\title{A Data-Driven Model Predictive Control Framework for Multi-Aircraft TMA Routing Under Travel Time Uncertainty} 
\author{Yi Zhang, Yushen Long, Liping Huang, Yicheng Zhang, Sheng Zhang, Yifang Yin
\thanks{Yi Zhang, Yushen Long, Liping Huang, Yicheng Zhang, Sheng Zhang and Yifang Yin are with the Institute for Infocomm Research (I$^2$R), Agency for Science, Technology and Research (A*STAR), 1 Fusionopolis Way, \#21-01 Connexis, Singapore 138632, Republic of Singapore. Emails: yzhang120@e.ntu.edu.sg (zhang$\textunderscore$yi@i2r.a-star.edu.sg), long$\textunderscore$yushen@i2r.a-star.edu.sg, huang$\textunderscore$liping@i2r.a-star.edu.sg, zhang$\textunderscore$yicheng@i2r.a-star.edu.sg, zhang$\textunderscore$sheng@i2r.a-star.edu.sg, yin$\textunderscore$yifang@i2r.a-star.edu.sg}%
\thanks{This work is supported by the National Research Foundation, Singapore, and the Civil Aviation Authority of Singapore (CAAS), under the Aviation Transformation Programme. (Grant No. ATP$\textunderscore$IOP for ATM$\textunderscore$I2R$\textunderscore$2)}
}
\maketitle
\begin{abstract}
This paper presents a closed-loop framework for conflict-free routing and scheduling of multi-aircraft in Terminal Manoeuvring Areas (TMA), aimed at reducing congestion and enhancing landing efficiency. Leveraging data-driven arrival inputs (either historical or predicted), we formulate a mixed-integer optimization model for real-time control, incorporating an extended TMA network spanning a 50-nautical-mile radius around Changi Airport. The model enforces safety separation, speed adjustments, and holding time constraints while maximizing runway throughput. A rolling-horizon Model Predictive Control (MPC) strategy enables closed-loop integration with a traffic simulator, dynamically updating commands based on real-time system states and predictions. Computational efficiency is validated across diverse traffic scenarios, demonstrating a 7-fold reduction in computation time during peak congestion compared to one-time optimization, using Singapore’s ADS-B dataset. Monte Carlo simulations under travel time disturbances further confirm the framework’s robustness. Results highlight the approach’s operational resilience and computational scalability, offering actionable decision support for Air Traffic Controller Officers (ATCOs) through real-time optimization and adaptive replanning.
\end{abstract}
\section{Introduction}

Flight delays impose high costs on passengers, airlines, and the environment due to increased fuel usage. As reported by Changi Airport \cite{Traffic_Statistics}, commercial flight activity has nearly returned to pre-COVID levels, with ~30,000 monthly movements from May to July and passenger volumes surpassing 2019 figures in March. To handle growing traffic, ATCOs frequently resort to radar vectoring to manage the influx of arriving and departing flights, highlighting a major challenge in current airspace operations.

At the planning level, Air Traffic Flow Management (ATFM) is typically modeled as a dynamic multi-commodity network-flow problem \cite{sun2008multicommodity}, incorporating strategies like ground holding, airborne holding, and rerouting to manage traffic across multiple airports without exceeding capacity limits \cite{zhang2018hierarchical}. However, within the Terminal Manoeuvring Area (TMA)—a high-density airspace near the airport—most literature primarily addresses Aircraft Scheduling Problems (ASP) for runway operations, with limited focus on optimizing the overall TMA airspace. The widely cited mixed-integer linear programming (MILP) model by Beasley et al. \cite{beasley2000scheduling} minimizes landing time deviations across multiple runways. Subsequent work extends this to interdependent runways, integrating departure scheduling and weather impacts \cite{lieder2016scheduling, pohl2021runway}. Meta-heuristic methods have also been explored to improve computational performance \cite{hammouri2020isa, zhang2024study}.

On the other hand, research on TMA airspace optimization is typically an extension of runway scheduling. Murca et al. introduce a job shop scheduling model that includes route selection but only emphasize runway safety separation, while neglecting airspace waypoint separation \cite{murcca2015control}. Sama et al. adopt the concept of alternative arcs, allowing waypoint-level control for holding operations and airspace safety separation, but lacks speed recommendations and uses a simple TMA map with limited waypoints \cite{sama2014optimal}. A receding horizon structure is introduced in \cite{sama2013rolling}, later also replacing branch and bound algorithm with a tabu search evolutionary algorithm in \cite{sama2017metaheuristics}. Furthermore, an optimization model \cite{ng2024optimization} is proposed that considers runway selection, speed adjustment, holding control, and vectoring. However, the model uses a simplistic map that does not address routing problems, and the aircraft order is predetermined on the runway, leading to a fixed order in airspace and preventing global optimization. A directed acyclic graph model for TMA multi-aircraft scheduling is proposed in \cite{ng2021mathematical}, leveraging extended Benders decomposition to improve efficiency. While it handles waypoint-based deconflictions, it lacks edge-level anti-overtaking constraints. In \cite{saez2019automation}, a MILP model divides the map into grid segments, with sides and diagonals as route options. Unlike our approach, it discretizes arrival times at waypoints, requiring many binary variables for separation constraints, which increases computational load.

Although recent advancements have begun to address the sequencing and scheduling of TMA airspace, a notable gap remains in developing a comprehensive model that accounts for the complex extended TMA network of the entire Standard Terminal Arrival Routes (STARs). This paper presents a closed-loop MPC framework for TMA traffic management, integrating a high-fidelity optimization model with a dynamic simulator to enable real-time conflict-free routing and scheduling. Unlike prior works limited to simplified networks, our approach models the entire Standard Terminal Arrival Route (STAR) structure within a 50-nautical-mile radius of Changi Airport, optimizing route selection, speed adjustments, and holding times under safety constraints to maximize runway throughput. Key innovations include:
\begin{itemize}
\item Data-driven inputs: Leveraging ADS-B data and an XGBoost predictor, we derive reliable Estimated Arrival Times (ETAs) at the TMA boundary as optimization inputs, bypassing error-prone runway ETA predictions.

\item Real-time MPC integration: A novel mixed-integer linear programming (MILP) model interacts bidirectionally with a simulator via rolling-horizon MPC, dynamically updating schedules using live traffic states and disturbances.

\item Scalable closed-loop validation: Comparative tests against one-time optimization show a 7× computational speedup during peak congestion (36 landing aircraft/hour), while Monte Carlo simulations under travel time uncertainty confirm robustness with high feasibility retention compared to Dijkstra’s algorithm. 
\end{itemize}

The rest of this paper is organized as follows. Section \ref{sec:framework} introduces the framework of whole system, where the system data-driven inputs are explained in Section \ref{sec:data}, the conflict-free optimization model in Section \ref{sec:model}, and the operation of the closed-loop simulation in Section \ref{sec:simulator}. In Section \ref{sec:experiment}, experiments on various case studies are conducted to illustrate computation efficiency and robustness. The experiment results and animation video can be found from \cite{zy2025github}. Finally, conclusion is summarized in Section \ref{sec:conclusion}.

\section{System Framework} \label{sec:framework}
\begin{figure*}[!ht]
	\centering
	\includegraphics[width=5.5in]{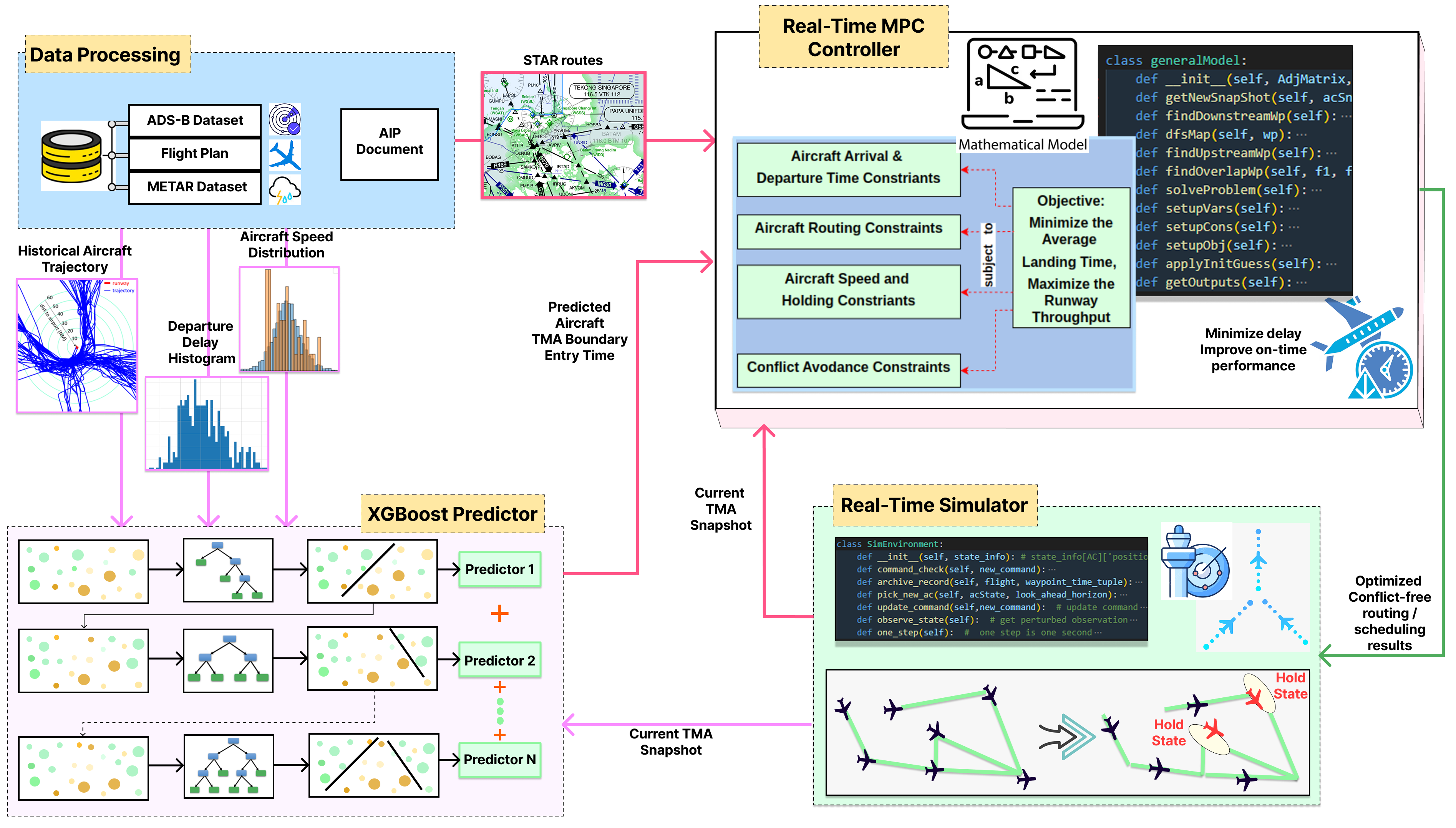}
	\caption{Schematic of a closed-loop system framework}
	\label{fig_frame}
\end{figure*}

The arrival scheduling system (Fig. \ref{fig_frame}) integrates four components: a historical database, a traffic predictor, a real-time MPC controller and a custom traffic simulator. Input data includes aircraft trajectories (ADS-B radar data),  flight plans and weather data. Machine learning method extracts traffic patterns from these datasets. The MPC optimizes aircraft scheduling (landing times, routes, speeds) via a noval MILP model, minimizing average landing times. Using real-time simulator outputs and predicted traffic flows, MPC commands are applied in a rolling horizon framework: optimized controls are executed in the simulator, which updates system states iteratively until simulation completion. The controller’s optimization cycle restarts with updated simulator snapshots and predictor forecasts. Section \ref{sec:data} summarizes data processing and predictor; Section \ref{sec:model} details the optimization formulation, and Section \ref{sec:simulator} explains closed-loop integration with the simulator.

\subsection{Data analysis on aircraft TMA boundary arrival time} \label{sec:data}
\begin{figure}[H]
	\centering
	\includegraphics[width=2.5in]{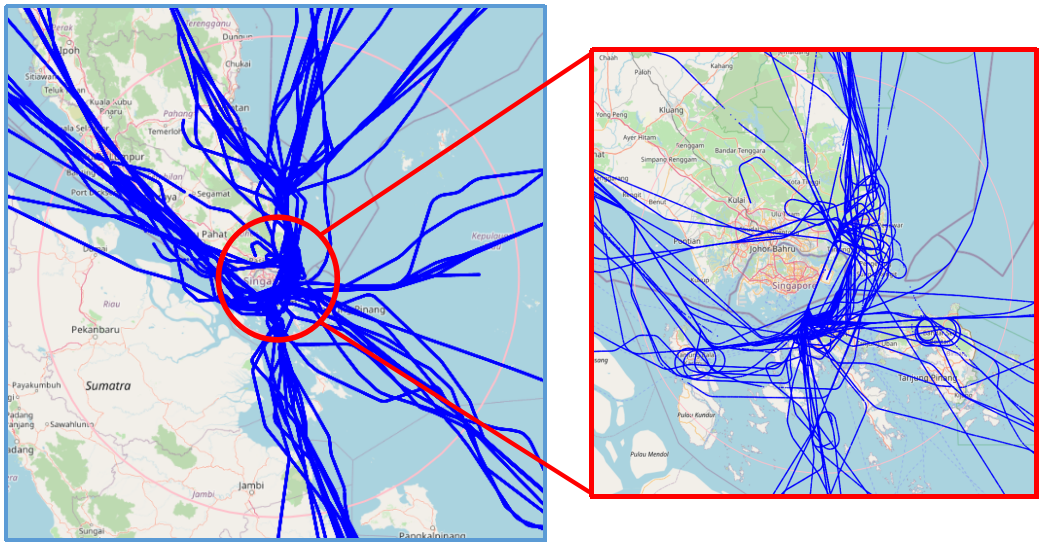}
	\caption{Trajectory Data: left panel (within 200 nautical mile), right panel (within 50 nautical mile) }
	\label{fig_trajData}
\end{figure}
Fig. \ref{fig_trajData} depicts one day aircraft trajectories from the ADS-B dataset. The left panel displays a wide-range view, encompassing a 200-nautical-mile (nm) radius from Changi airport, while the right panel provides a zoomed-in view of the inner circle, representing the studied TMA approximately 50 nm from Changi airport. Significant holding and vectoring patterns are distinctly observed within the inner circle.

\begin{figure}[H]
	\centering
	\includegraphics[width=3.0in]{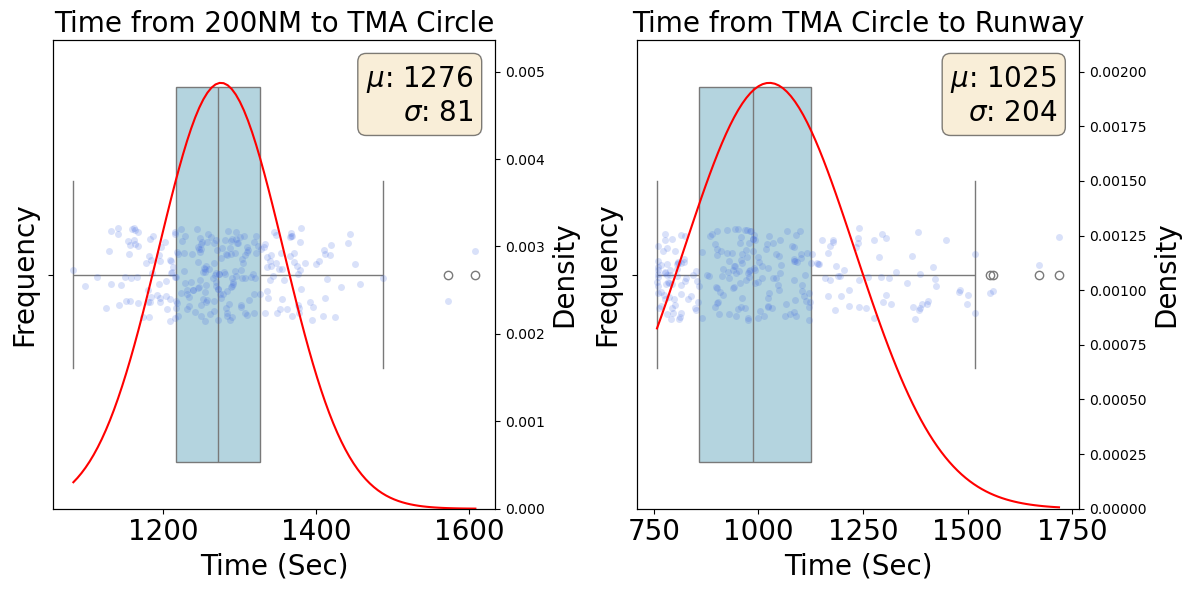}
	\caption{Data analysis on travel time distribution}
	\label{fig_distributionData}
\end{figure}

This can be further verified in Fig. \ref{fig_distributionData}, which presents a box plot of travel times and their corresponding fitted normal distributions. The left panel represents data for the larger range, extending 200 nm toward the TMA boundary, while the right panel focuses on the area from the TMA boundary to the final runway. Clearly, The inner TMA area exhibits a significantly larger variance (204s) compared to the studied enroute area (81s). Furthermore, despite the enroute segment covering approximately three times the distance of the inner TMA, the average travel time $\mu$ for both regions are quite close. This highlights higher aircraft speeds in the enroute area and the frequent occurrence of holding and vectoring operations within the TMA.

We propose to predict the 50NM boundary arrival time for each inbound aircraft. The truth label for aircraft $f$ at a given timestamp $t$ is $\mathbf{T}_f^t$, which is calculated as $\mathbf{T}_f^t=\mathbf{t}_{f}^{B} - t$, where $t$ is the given timestamp at which the arrival time prediction for aircraft $f$ is conducted, and $\mathbf{t}^{B}_{f}$ is the truth timestamp when aircraft $f$ arrives at the 50NM boundary, extracted from the historical track data. 
For predicting $\mathbf{T}_f^t$, we construct a mapping function $\textbf{T}_f^t=\mathcal{F}(X_f^t)$, where $X_f^t$ is the feature inputs that capture the factors of boundary arrival time modeling at time $t$ for aircraft $f$. In this study, we adopt XGBoost \cite{chen2016xgboost} to construct the mapping function, and the input feature $X_f^{t}$ is a combination of positioning, temporal and meteorological information. The objective is to construct the mapping function as 

\begin{equation}
    \mathbf{T}_{f}^t = \mathcal{F}(X_{f}^t), f\in\mathcal{J}_t, t\in\mathcal{T}
    \label{target}
\end{equation}
where $\mathcal{J}_t$ is the set of all aircraft in the dataset in the training phase. $\mathcal{T}$ is all the timestamp set, e.g., for a one-day dateset and data sampling is one point per second, then timestamp set $\mathcal{T}$ is all seconds in this day. $X_{f}^t = [X_{f, dyn}^t, X_{rwy}^t, X_{f, ADEP}, X_{f, WTC}, X_{temp}^t, X_{METAR}^t]$. 

\begin{figure}[!ht]
	\centering
	\includegraphics[width=3.0in, height=1.3in]{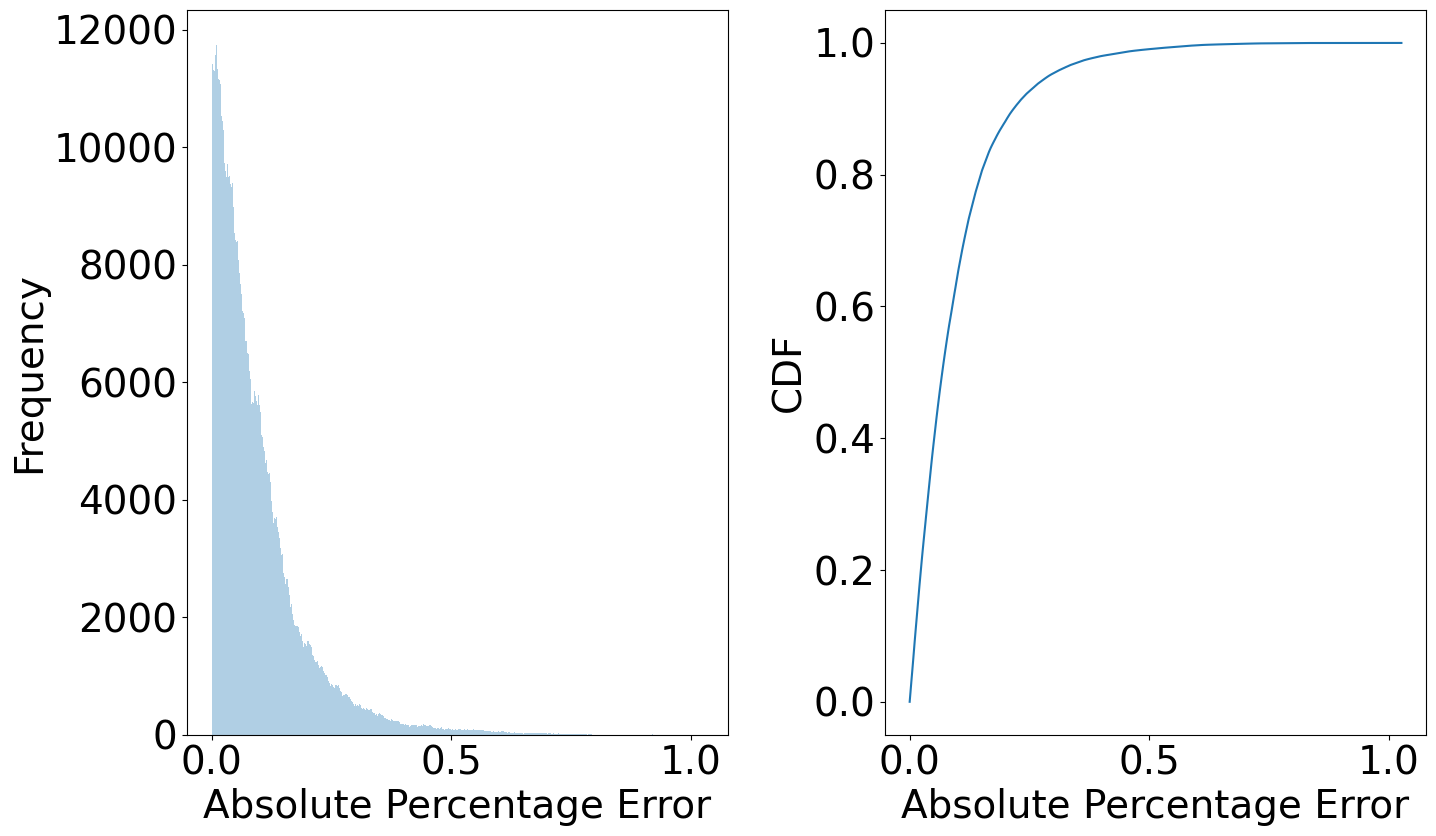}
	\caption{Prediction on TMA boundary arrival time}
	\label{fig_predict}
\end{figure}

  Specifically, features from ADS-B dataset and METeorological Aerodrome Report (METAR) data \cite{metarData} are utilized. The real-time dynamic information of aircraft $f$ at timestamp $t$, $X_{f, dyn}^t$, is composed of the latitude, longitude, ground speed and distance to Changi airport. $X_{rwy}^t$ is the runway direction at timestamp $t$, e.g., \enquote{02} (\enquote{20}) denotes south (north) landing in Changi airport. $X_{f, ADEP}$ is the departure airport ICAO code for aircraft $f$ in the flight plan, e.g., \enquote{WMKK} (Kuala Lumpur Airport). $X_{f, WTC}$ is the wake turbulence category of aircraft $f$, $X_{temp}^t$ is the temporal attributes (hour, day, and week), and $X_{METAR}^t$ is the meteorological features derived from METAR of the past latest 30min snapshot (METAR data granularity is 30 minutes), including the wind speed, wind direction, visibility, sky level 1 coverage and altitude in feet. Fig. \ref{fig_predict} illustrates the prediction result on one-day data, where the absolute error is 170.35s. The left panel illustrates the Absolute Percentage Error (APE) distribution, which demonstrates that most errors are concentrated near zero, indicating a high level of prediction accuracy. The right panel shows the cumulative distribution function (CDF) of APE, where the curve has a large slope and quickly increase to a value near 1.0, indicating the reliability of the model. These results confirm the model's overall robustness in predicting TMA boundary arrival times. 

Predicting the aircraft’s arrival time at the TMA boundary is more straightforward than directly forecasting the ETA on the runway, as holding patterns and speed adjustments are commonly applied within the TMA due to higher traffic density. This further validates the correctness of the current framework, which adopts the predicted boundary arrival time as inputs and then performs optimization within the TMA until landing.

\subsection{Optimization controller formulation} \label{sec:model}
\begin{figure}[!ht]
	\centering
	\includegraphics[width=2.0in]{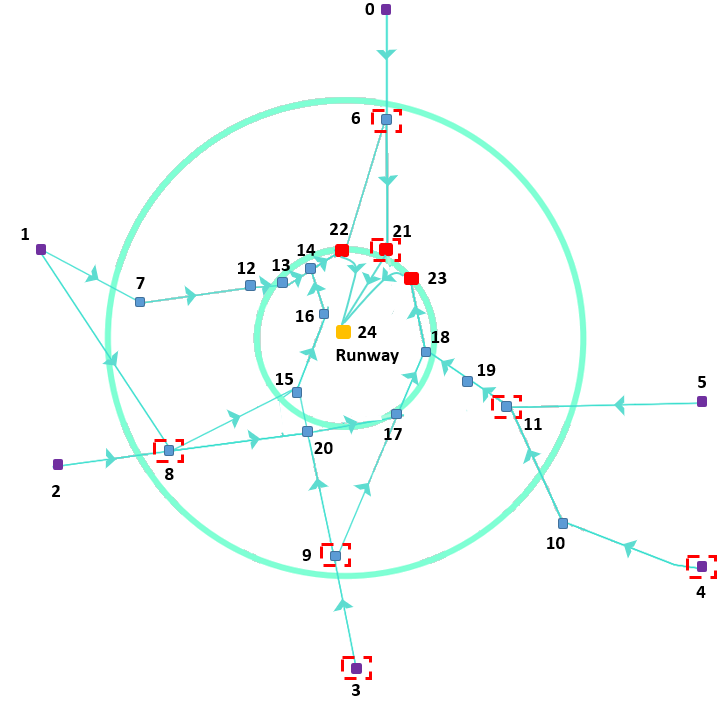}
	\caption{TMA Runway 20 STARs (Captured from AIP 2022)}
	\label{fig_route}
\end{figure}

The TMA airborne multi-aircraft control task must ensure conflict-free routes while optimizing for efficient landings. This decision-making process necessitates a long-horizon perspective from the current flight position to the final runway, rather than a sequential determination approach. Consequently, this problem is formulated as an Operation Research (OR)-driven challenge rather than a dynamic-based trajectory tracking problems. However, the optimized OR results shall be converted to time-indexed alternatives in the simulator to enable the time evolution in Section \ref{sec:simulator}. TMA is normally a potential bottleneck where flights from different directions all trying to land at the airport. To facilitate the large volume landings,  STARs are designed, as depicted in Fig. \ref{fig_route}, which illustrates the STARs map retrieved from Aeronautical Information Publication (AIP), depicting routes from TMA boundary towards landing runway 20 for Changi airport. The waypoints enclosed within the red dashed box are designated for holding maneuvers. The proposed model focus on 2D deconflict, and assume the landing process shall follow a continuous
descend to the next waypoint altitude as required in the AIP document. The following subsections formulate the problem as a mixed-logic problem, aiming to minimize the average landing time for all aircraft, incorporating constraints such as route conservation law, arrival and departure constraints, speed and holding constraints and conflict avoidance constraints.
\subsubsection{Source and sink constraints}
The start of the optimization needs to know when the aircraft approaches its immediately following waypoint, also, when rolling-horizon mechanism is activated, aircraft can approach any waypoints inside TMA at any time within look-ahead horizon. Thus, the initial assignment constraints shall be illustrated as follows:
\begin{subequations}\label{eq:assignment}
\allowdisplaybreaks
	\begin{align}
		& \forall f \in F, \nonumber\\
		& \sum_{j\in W} x_{s_{f}j}^{f} = 1\\
        & DP_{s_{f}}^{f} = DS_{s_{f}}^{f} + HT_{s_{f}}^{f}
	\end{align}
\end{subequations}
where $x_{ij}^{f}$ denotes whether link $ij$ is selected for aircraft $f$ from node $i$ to node $j$. $DP_{i}^{f}$ and $HT_{i}^{f}$ are the departure time and holding time of $f$ at node $i$, respectively. $DS_{i}^{f}$ is known parameter denoting the incoming time retrieved either from simulator or predictor. $W$ is the waypoint set in TMA map, $F$ is the aircraft set, $s_{f}$ is the initial waypoint of aircraft $f$, $s_{f} \in W$.

All arrival aircraft shall finally land on runways, as captured below:
\begin{equation}\label{eq:Sink}
\allowdisplaybreaks
	\begin{aligned}
        & \forall j \in E, \sum_{f\in F}\sum_{i\in W} x_{ij}^{f} = |F|
	\end{aligned}
\end{equation}
E denotes the runway set, and $|F|$ is the total number of aircraft within the current look-ahead horizon.
\subsubsection{Route conservation law}
Similar to the vehicle routing problem, the sum of routing variables $x_{ij}^{f}$ from all upstream nodes shall equal to the sum of $x_{jq}^{f}$ towards all downstream nodes. Also, there is no subtour elimination constraint as TMA map is an acyclic network.
\begin{equation}\label{eq:conservation1}
\allowdisplaybreaks
	\begin{aligned}
		& \forall j \in W, \forall f \in F, \sum_{i \in W, i \neq j} x_{ij}^{f} = \sum_{q \in W, q \neq j} x_{jq}^{f} 
	\end{aligned}
\end{equation}
The connection of the route strictly follow the TMA map, only if the link $ij$ in adjacency matrix of TMA map $A_{ij}$ exists, then route selection is possible.
\begin{equation} \label{eq:adj}
	\forall i, j \in W, \forall f \in F, x_{ij}^{f} \leq A_{ij}
\end{equation}
\subsubsection{Arrival and departure constraints}
When aircraft $f$ select link $ij$, then its arrival time at node $j$ shall equal to the sum of its departure time at $i$ and travel time on link $ij$.
\begin{subequations} \label{eq:arrDep}
\allowdisplaybreaks
	\begin{align}
		& \forall i, j \in W, \forall f \in F \nonumber\\
		& AR_{j}^{f} - DP_{i}^{f} - d_{ij}\sum_{s}(\hat{v}_{ijs}\delta_{ijs}^{f}) \leq M(1-x_{ij}^{f})  \\
        & -AR_{j}^{f} + DP_{i}^{f} + d_{ij}\sum_{s}(\hat{v}_{ijs}\delta_{ijs}^{f}) \geq - M(1-x_{ij}^{f}) \\
        &\sum_{s}\delta_{ijs}^{f} = 1
	\end{align}
\end{subequations}
where $\hat{v}_{ijs}$ indicates the reciprocal of the speed in link $ij$ at level $s$, $d_{ij}$ is the distance between nodes $i$ and $j$, $\delta_{ijs}^{f}$ is the binary variable used to select the appropriate speed level for link $ij$, $M$ is a big value. The specified speed levels are provided due to the command-based communication between pilots and air traffic control officers, where speed commands must adhere to predetermined values. Also, each time only one speed level can be selected.

On the other hand, the departure time at $i$ (excluding the initial node) is the sum of the arrival time at $i$ and the possible holding time.
\begin{subequations} \label{eq:depArr}
\allowdisplaybreaks
	\begin{align}
		& \forall f \in F, \forall i \in W \setminus s_{f} \nonumber \\
		& DP_{i}^{f} = AR_{i}^{f} + HT_{i}^{f} \\
        & HT_{i}^{f} \leq M_{h} H_{i}
	\end{align}
\end{subequations}
where $M_h$ is the maximum holding time, and $H_{i}$ denotes whether node $i$ allows holding or not. 
\subsubsection{Safety gap constraints}
The ICAO manual \cite{ICAO_Global_Aviation_Safety_Plan_2022} defines the safety regulation that a certain gap must be guaranteed for any two aircraft inside TMA. Thus, we incorporate this restriction as following logic constraints:
\begin{subequations}\label{eq:conflict}
\allowdisplaybreaks
	\begin{align}
		& \forall i, j \in W, \forall f, f^{'} \in F, \nonumber \\
		& \sum_{i}x_{ij}^{f} = 1 \wedge \sum_{i}x_{ij}^{f^{'}} = 1 \rightarrow \nonumber \\
  &  (AR_{j}^{f} \geq AR_{j}^{f^{'}}+t_{f,f^{'}}) \cup (AR_{j}^{f}+t_{f,f^{'}}\leq AR_{j}^{f^{'}}) \\
  & \sum_{i}x_{ij}^{f} = 1 \wedge \sum_{i}x_{ij}^{f^{'}} = 1 \rightarrow \nonumber \\
  &  (DP_{j}^{f} \geq DP_{j}^{f^{'}}+t_{f,f^{'}}) \cup (DP_{j}^{f}+t_{f,f^{'}}\leq DP_{j}^{f^{'}})
	\end{align}
\end{subequations}
For any two aircraft, $f$ and $f^{'}$, if both reaching waypoint $j$, either aircraft $f$ arrives first or aircraft $f^{'}$ does, but ensuring a minimum gap $t_{f,f^{'}}$ between them is imperative, similar to departure operation.
\subsubsection{Rear-end constraints}
\begin{figure}[!ht]
	\centering
	\includegraphics[width=3.0in]{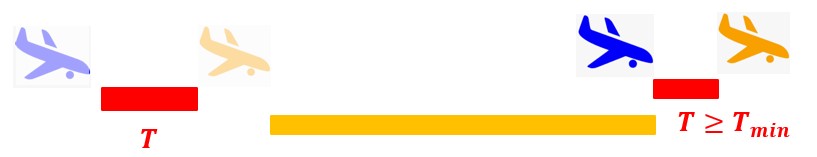}
	\caption{Schematic of rear-end constraint}
	\label{fig_overtake}
\end{figure}

As depicted in Fig. \ref{fig_overtake}, overtaking is prohibited on the link, following the First-Come-First-Serve principle typically observed by ATCOs in TMAs. Additionally, we assume speed selection pertains to the entire link, precluding speed changes within it. Therefore, our focus lies on ensuring a maintained safety gap between two aircraft upon completing travel at the downstream node.

To address this, the following logic constraints are introduced: If two aircraft, $f$ and $f^{'}$, both choose link $ij$, and the departure time of $f$ at waypoint $i$ exceeds that of $f^{'}$, then the arrival time of $f$ at waypoint $j$ surpasses that of $f^{'}$ by at least the minimum safety gap $t_{f,f^{'}}$.
\begin{equation}\label{eq:anti-overtake}
\allowdisplaybreaks
	\begin{aligned}
		& \forall i, j \in W, x_{ij}^{f} = 1 \wedge x_{ij}^{f^{'}} = 1 \wedge DP_{i}^{f}>DP_{i}^{f^{'}}\nonumber\\
		& \rightarrow AR_{j}^{f}\geq AR_{j}^{f^{'}} + t_{f,f^{'}}
	\end{aligned}
\end{equation}
\subsubsection{Objective}
In this model, our primary objective is to minimize the average landing time of all aircraft, thereby enhancing overall operational efficiency within the TMA airspace. This optimization aims to expedite the arrival process and reduce congestion, leading to improved airspace utilization and enhanced runway throughput. The objective is illustrated as follows:
\begin{equation}\label{eq:cost}
	J = \text{min} \frac{1}{|F|}\underset{f \in F, j \in E}{\sum} AR_{j}^{f} 
\end{equation}

\subsection{Closed-loop Formulation} \label{sec:simulator}
The state of the studied traffic network system at time instant $k$ is defined as $y(k)=\{(\text{wp}_f(k), t_{\text{arr},f}(\text{wp}_f(k)), t_{\text{hld},f}(\text{wp}_f(k))),f\in F\}$, where $F$ is the set of flight, $f$ is the index of flights, $\text{wp}_f(k)$ is the next waypoint of flight $f$ at time instant $k$, $t_{\text{arr},f}(\text{wp}_f(k))$ is the remaining time for flight $f$ to reach $\text{wp}_f(k)$ at time instant $k$ and $t_{\text{hld},f}(\text{wp}_f(k)))$ is the remaining holding time at $\text{wp}_f(k)$. We use the convention that the next waypoint is updated only after flight $f$ leaves current waypoint.

The flight command from optimizer is essentially a function of $y_f(k)$, which we denote as $u_f^{opt}(k)$. We also denote the flight command applied to flights as $u_f(k)$. Once a new solution from the optimizer is sent to the set of flights, the existing command $u_f(k)$ will be set as $u_f^{opt}(k)$. For each flight, the command is a sequence of waypoints to be visited, a sequence of remaining time to reach each waypoint, and holding time at each waypoint. In particular, at time instant $k$, the command for flight $f$ shall be $u_f(k)=(S^k_{\text{wp},f},S^k_{\text{arr},f},S^k_{\text{hld},f})$, where $S^k_{\text{wp},f}=(\text{wp}^u_{f,1}(k),\ldots,\text{wp}^u_{f,N}(k))$, $S^k_{\text{arr},f}=(t^u_{\text{arr},f}(\text{wp}_{f,1}(k)),\ldots,t^u_{\text{arr},f}(\text{wp}_{f,N}(k)))$, and $S^k_{\text{hld},f}=(t^u_{\text{hld},f}(\text{wp}_{f,1}(k)),\ldots,t^u_{\text{hld},f}(\text{wp}_{f,N}(k)))$.
\subsubsection{Nominal Case}
In this case, we assume that there is no uncertainty in the studied system. For each flight $f$, its state $y_f(k)=(\text{wp}_f(k), t_{\text{arr},f}(\text{wp}_f(k)), t_{\text{hld},f}(\text{wp}_f(k)))$ and command $u_f(k)=(S^k_{\text{wp},f},S^k_{\text{arr},f},S^k_{\text{hld},f})$ are updated as follows:

\textbf{Case 1)}: $t_{\text{arr},f}(\text{wp}_{f,1}(k))\ge1$,
\begin{equation}
\allowdisplaybreaks
\begin{aligned}
    \text{wp}_{f}(k+1)&=\text{wp}_{f}(k),\\
    t_{\text{arr},f}(\text{wp}_{f,1}(k+1))&=t_{\text{arr},f}(\text{wp}_{f,1}(k))-1,\\
    t_{\text{hld},f}(\text{wp}_f(k+1))&=t_{\text{hld},f}(\text{wp}_f(k)),\\
    S^{k+1}_{\text{wp},f}&=S^k_{\text{wp},f},\\
    S^{k+1}_{\text{arr},f}&=(t-1~\text{for}~t~\text{in }S^k_{\text{arr},f}),\\
    S^{k+1}_{\text{hld},f}&=S^k_{\text{hld},f}. \nonumber
\end{aligned}
\end{equation}

In case 1), flight is still on the way to next waypoint, and thus the next waypoint and holding time remain the same, while all remaining arrival time shall deduct 1 time unit.

\textbf{Case 2)}: $t_{\text{arr},f}(\text{wp}_{f,1}(k))<1$ and $t_{\text{hld},f}(\text{wp}_f(k))\ge1-t_{\text{arr},f}(\text{wp}_{f,1}(k))$. 
\begin{equation}
\allowdisplaybreaks
\begin{aligned}
    \text{wp}_{f}(k+1)&=\text{wp}_{f}(k),\\
    t_{\text{arr},f}(\text{wp}_{f,1}(k+1))&=0,\\
    t_{\text{hld},f}(\text{wp}_f(k+1))&=t_{\text{hld},f}(\text{wp}_f(k))-(1-t_{\text{arr},f}(\text{wp}_{f,1}(k))),\\
    S^{k+1}_{\text{wp},f}&=S^k_{\text{wp},f},\\
    S^{k+1}_{\text{arr},f}&=(t-1~\text{for}~t~\text{in }S^k_{\text{arr},f}),\\
    S^{k+1}_{\text{hld},f}&=S^k_{\text{hld},f}.\nonumber
\end{aligned}
\end{equation}

In case 2), flight will arrive at the next waypoint and hold there. Therefore, by our convention that waypoint is updated only after leaving, $\text{wp}_{f}(k+1)$ remains the same and the remaining arrival time of this waypoint is set as 0. The remaining holding time at this waypoint and arrival time of future waypoints are updated accordingly.

\textbf{Case 3)}: $t_{\text{arr},f}(\text{wp}_{f,1}(k))<1$ and $t_{\text{hld},f}(\text{wp}_f(k))<1-t_{\text{arr},f}(\text{wp}_{f,1}(k))$.
\begin{equation}
\allowdisplaybreaks
\begin{aligned}
    \text{wp}_f(k+1)&=\text{wp}_{f,2}^u(k),\\
    t_{\text{arr},f}(\text{wp}_{f,1}(k+1))&=t^u_{\text{arr},f}(\text{wp}_{f,2}(k))-1,\\
    t_{\text{hld},f}(\text{wp}_f(k+1))&=t^u_{\text{hld},f}(\text{wp}_{f,2}(k)),\\
    S^{k+1}_{\text{wp},f}&=\text{pop}(S^k_{\text{wp},f}),\\
    S^{k+1}_{\text{arr},f}&=\text{pop}((t-1~\text{for}~t~\text{in }S^k_{\text{arr},f})),\\
    S^{k+1}_{\text{hld},f}&=\text{pop}(S^k_{\text{hld},f}).\nonumber
\end{aligned}
\end{equation}

where $\text{pop}(\cdot)$ denotes the operation of removing the first element of a sequence and relabeling rest elements. 

In case 3), flight will leave current waypoint, so arrival time and holding time of the next waypoint are updated from the command list. We also remove the first command in the command list as it has been executed.

\subsubsection{Disturbance Case}\label{sec:disturbexp3}
In this case, we introduce two uncertainties in state update and state observation. The uncertainty for state update aims to simulate scenarios when flights do not strictly follow the time schedule due to possible weather conditions etc. The uncertainty from state observation is corresponding to the fact that we cannot precisely predict the arrival time of next waypoint of each flight. To introduce the first uncertainty, we shall revise Case 3) above as follows:

\textbf{Case 3)}:
\begin{equation}
\allowdisplaybreaks
\begin{aligned}
    \text{wp}_f(k+1)&=\text{wp}_{f,2}^u(k),\\
    t_{\text{arr},f}(\text{wp}_{f,1}(k+1))&=(1+\xi_{\text{arr},f})(t^u_{\text{arr},f}(\text{wp}_{f,2}(k))-1),\\
    t_{\text{hld},f}(\text{wp}_f(k+1))&=(1+\xi_{\text{hld},f})t^u_{\text{hld},f}(\text{wp}_{f,2}(k)),\\
    S^{k+1}_{\text{wp},f}&=\text{pop}(S^k_{\text{wp},f}.\text{pop}),\\
    S^{k+1}_{\text{arr},f}&=\text{pop}((t-1~\text{for}~t~\text{in }S^k_{\text{arr},f})),\\
    S^{k+1}_{\text{hld},f}&=\text{pop}(S^k_{\text{hld},f}).\nonumber
\end{aligned}
\end{equation}
where $\xi_{\text{arr},f}$ and $\xi_{\text{hld},f}$ are two random variables with desired distribution.

Besides the uncertainty involved in the state update, the second uncertainty is also applied to state observation when passing to the optimizer. The state to be fed to the optimizer in this case is not the real state $y_f(k)$ but an estimation with uncertainty $\tilde{y}_f(k)=(\text{wp}_f(k), \tilde{t}_{\text{arr},f}(\text{wp}_f(k)), \tilde{t}_{\text{hld},f}(\text{wp}_f(k)))$, where $\tilde{t}_{\text{arr},f}(\text{wp}_f(k))=(1+\xi_{ob,f})t_{\text{arr},f}(\text{wp}_f(k))$, $\tilde{t}_{\text{hld},f}(\text{wp}_f(k))=(1+\xi_{ob,f}t_{\text{hld},f}(\text{wp}_f(k)))$ and $\xi_{ob,f}$ is a random variable with desired distribution. In this case we assume that the next waypoint is determined but remaining arrival time and holding time are not known precisely. 

\section{Experiment Results}\label{sec:experiment}
\subsection{Computational Complexity}
The problem described in Section \ref{sec:model} encompasses logic constraints, which are transformed into a MILP problem using the big-M method. Subsequently, this MILP problem is implemented in Python and solved utilizing the commercial solver Gurobi \cite{gurobi2015gurobi}. The programming code is running on a PC with 13th Gen Intel Core i9-13900KF × 32 CPU up to 5.80 GHz and RAM 31GB. To investigate the proposed model, we have summarized 9 cases under different number of aircraft, as illustrated in Table \ref{tab:cases}, where $N^f$, $N_{cons}$ and $N_{vars}$ denote the number of aircraft, constraints, and variables, respectively. $gap_{min}$ represents the minimum gap (unit is second) for any two aircraft approaching the same initial waypoint, the whole testing requires the safe gap $t_{f,f^{'}}$ is 60s, in other words, if two aircraft arrive at initial waypoint with 30s gap, their safety gap with any other aircraft (including each other) must be increased to at least one minute at the following subsequent waypoints, which becomes more challenging. The experiments in this subsection only focus on the complexity study of the optimization model, with no connection to the simulator component. Although the number of aircraft varies from 10 to 27 across the 9 cases, all aircraft arrive at the TMA boundary within a 10-minute window, creating a level of congestion surpassing typical real-world scenarios. For historical data, the most congested landing events generally involve no more than 40 aircraft per hour. In contrast, our tests involve scenarios with 20+ landing aircraft within just 10 minutes, reflecting an even higher level of congestion. This is designed to further assess the solver's computational capacity.
\begin{table}[!h]
\caption{Test cases under different parameter settings}
\begin{center} \label{tab:cases}
\resizebox{0.8\linewidth}{!}{
\begin{tabular}{c|c|c|c|c}
\hline
\hline
Cases & {\begin{tabular}[c]{@{}c@{}}$N_{f}$\\ Arrivals within 10 min\end{tabular}}  & $gap_{min}$ & $N_{cons}$ & $N_{vars}$ \\ \hline
Case 1  & 10 & 60 & 36001 & 43820              \\ \hline
Case 2  & 12 & 60 & 44571 & 53592           \\ \hline
Case 3  & 15 & 60 & 59741 & 68880             \\ \hline
Case 4  & 10 & 30 & 36711 & 43820             \\ \hline
Case 5  & 12 & 30 & 45587 & 53592     \\ \hline
Case 6  & 15 & 30 & 59819 & 68880    \\ \hline
Case 7  & 23 & 60 & 106657 & 113344            \\ \hline
Case 8  & 25 & 30 & 118271 & 125300     \\ \hline
Case 9  & 27 & 30 & 134259 & 137592         \\ \hline
\end{tabular}
}
\end{center}
\end{table}

The problem scale increases with the increase of the aircraft number, as depicted in Table \ref{tab:cases}, the corresponding MILP results are listed in Table \ref{tab:computationTime}. Besides the proposed MILP formulation, we also investigate the Dijkstra algorithm, which has been broadly adopted to solve the path finding problem. The Dijkstra algorithm is sequentially applied to each aircraft, and the routes determined for all preceding aircraft are imposed as constraints for subsequent aircraft when computing their routes. This sequential approach enables the determination of a feasible solution after processing all aircraft. The order of aircraft visitation is determined by their respective ideal landing times, which are calculated as the sum of initial arrival time and ideal travel time.

\begin{table*}[!h]
\caption{Computational times and objective costs for 9 cases between MILP solver and Dijkstra algorithm}
\begin{center} \label{tab:computationTime}
\resizebox{\linewidth}{!}{
\begin{tabular}{c|c|c|cc|ccc|c}
\hline
\multirow{2}{*}{\textbf{}} & \multirow{2}{*}{\textbf{$N_{cons}$}} & \multirow{2}{*}{\textbf{$N_{vars}$}} & \multicolumn{2}{c|}{\textbf{MILP Solver}}                                                                                                                                  & \multicolumn{3}{c|}{\textbf{Priority-based Dijkstra Algorithm}}                                                                                                                                                                                                                            & \multirow{2}{*}{\textbf{Cost Gap}} \\ \cline{4-8}
                           &                                   &                                   & \multicolumn{1}{c|}{\textbf{\begin{tabular}[c]{@{}c@{}}Computational\\ Time (s)\end{tabular}}} & \textbf{\begin{tabular}[c]{@{}c@{}}Objective Cost \\of (\ref{eq:cost}) (s)\end{tabular}} & \multicolumn{1}{c|}{\textbf{\begin{tabular}[c]{@{}c@{}}Computational\\ Time (s)\end{tabular}}} & \multicolumn{1}{c|}{\textbf{\begin{tabular}[c]{@{}c@{}}Objective Cost \\of (\ref{eq:cost}) (s)\end{tabular}}} & \textbf{\begin{tabular}[c]{@{}c@{}}Infeasible Aircraft\\ Number\end{tabular}} &                   \\ \hline
\textbf{Case 1}            & 36001 & 43820 & \multicolumn{1}{c|}{2.3772} &1189.2 & \multicolumn{1}{c|}{14.990} & \multicolumn{1}{c|}{1191.5}  & 0  &  $0.1934\%$                 \\ \hline
\textbf{Case 2}            & 44571 & 53592 & \multicolumn{1}{c|}{3.0409} &1101.5 & \multicolumn{1}{c|}{15.470} & \multicolumn{1}{c|}{1103.4}  & 0  &  $0.1725\%$                 \\ \hline
\textbf{Case 3}            & 59741 & 68880 & \multicolumn{1}{c|}{4.8222} &1127.5 & \multicolumn{1}{c|}{18.985} & \multicolumn{1}{c|}{1129.0}  & 0  &  $0.1330\%$                \\ \hline
\textbf{Case 4}            & 36711 & 43820 & \multicolumn{1}{c|}{3.2988} &1155.6 & \multicolumn{1}{c|}{11.068} & \multicolumn{1}{c|}{1195.6}  & 0  &  $3.4614\%$                 \\ \hline
\textbf{Case 5}            & 45587 & 53592 & \multicolumn{1}{c|}{8.3390} &1104.4 & \multicolumn{1}{c|}{11.861} & \multicolumn{1}{c|}{1139.8}  & 0  &  $3.2054\%$                \\ \hline
\textbf{Case 6}            & 59819 & 68880 & \multicolumn{1}{c|}{20.499} &1084.7 & \multicolumn{1}{c|}{21.966} & \multicolumn{1}{c|}{1103.8}  & 0  &  $1.7609\%$                 \\ \hline
\textbf{Case 7}          & 106657 & 113344 & \multicolumn{1}{c|}{72.071} &1119.0 & \multicolumn{1}{c|}{24.031} & \multicolumn{1}{c|}{1121.6}  & 0 & $0.2324\%$                  \\ \hline
\textbf{Case 8}          & 118271 & 125300 & \multicolumn{1}{c|}{173.51} &1072.5 & \multicolumn{1}{c|}{24.481} & \multicolumn{1}{c|}{$\sim$1080.6}  & 2 & $\ast$                  \\ \hline
\textbf{Case 9}          & 134259 & 137592 & \multicolumn{1}{c|}{2609.6} &1083.9 & \multicolumn{1}{c|}{24.815} & \multicolumn{1}{c|}{$\sim$1085.3}  & 2 & $\ast$                 \\ \hline
\multicolumn{3}{l}{$\sim$ is the objective cost excluding infeasible aircraft.} \\
\multicolumn{3}{l}{$\ast$ means infeasibility leads to meaningless gap calculation.} 
\end{tabular}
}
\end{center}
\end{table*}

Clearly, the computational time exponentially increases as the increase of the problem scale for MILP solver: Case 9 is approximately three times larger than Case 1 in scale, yet the computation time for Case 9 is already 200 times greater than that of Case 1. Also, cases 4, 5, and 6 exhibit a similar scale to cases 1, 2, and 3, respectively. However, owing to the smaller value of $gap_{min}$ in cases 4, 5, and 6, their solutions necessitate more time to be found compared to cases 1, 2, and 3. On the other hand, the priority-based Dijkstra algorithm demonstrates rapid computation time, highlighting its efficient search strategy compared to the branch-and-bound algorithm to find global optimal values in the solver. Furthermore, its computation time increases polynomially with the expansion of the problem scale. Even though it follows a greedy way to find solution, the gap between its objective cost and optimal cost from MILP remains within a constrained range, consistently below 5$\%$. However, with the increase of the problem scale, Dijkstra starts to struggle in ensuring feasibility, leading to instances where certain aircraft fail to find a feasible route. This is attributed to its biased and greedy approach to solution finding, which narrows down the actual feasible region.

\subsection{Rolling-horizon based one-hour historical test} \label{sec:exp2}
\begin{figure}[!h]
	\centering
	\includegraphics[width=3.0in]{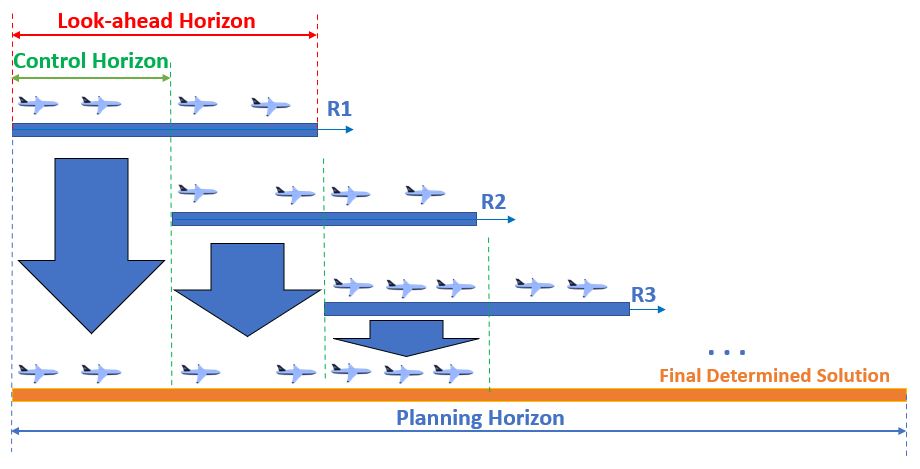}
	\caption{Schematic of rolling-horizon mechanism}
	\label{fig_rolling}
\end{figure}

To ensure both rapid computation and feasibility, we consider employing a rolling horizon approach for MILP to address the problem. Fig. \ref{fig_rolling} illustrates how MPC is conducted in a rolling-horizon fashion, where the whole planning horizon is partitioned into a group of subproblems to speed up the computation. Every time the calculation only happens in one look-ahead horizon, but only optimized results in control horizon shall be applied to the simulator and recorded as final determined solutions, the reason to include additional period in look-ahead horizon is to avoid shortsightedness when making decisions for aircraft in control horizon. 

To validate the proposed method, we have chosen various busy hours from the ADS-B dataset. Following data processing, we have derived multiple hourly scenarios representing the sequence of aircraft arrivals into the TMA. Table \ref{tab:exp2} lists out the simulation results with and without adopting rolling-horizon approach. The look-ahead horizon and control horizon are 10min and 5min, respectively. Clearly, as the problem scale increases, the computation time of one-time MILP exponentially increases. Also, not only the number of aircraft, but also different arrival aircraft distributions will lead to different computation times, as shown in Cases 2 and 3 in Table \ref{tab:exp2}. Moreover, Case 3 in Table \ref{tab:exp2}, comparable in scale to Case 9 in Table \ref{tab:cases}, exhibit significantly faster processing times. This is because the first experiment's cases are randomly generated within a 10-minute interval, whereas the current cases in Table \ref{tab:exp2} represent the number of aircraft from historical records spanning one hour, and already selected from busy hours. This observation suggests that the workload of ATCOs already constrains the potential additional capacity that could be accommodated in the TMA. Furthermore, through the implementation of a rolling-horizon approach, while we calculate the optimal solution for each look-ahead horizon, we observe minimal alteration in the final cost value (only slight increase on Case 2) compared to a one-time optimization. Nonetheless, significant reductions in computational time are evident. 
For a more comprehensive aircraft movement analysis of the results, the corresponding video is provided in the Experiment 2 folder of \cite{zy2025github}. In the video, the animation feature is enabled (note that the current demo implementation operates on a single thread) and saved as an MP4 file at its original speed, without any acceleration or deceleration. This results in higher time consumption compared to the results presented in Table \ref{tab:exp2}, where the animation is disabled.
\begin{table}[!h]
\caption{Comparison of MILP between the one-time optimization and rolling-horizon approach}
\begin{center} \label{tab:exp2}
\resizebox{0.95\linewidth}{!}{
\begin{tabular}{c|c|cc|cc}
\hline
\multirow{2}{*}{} & \multirow{2}{*}{$N^f$} & \multicolumn{2}{c|}{\textbf{One-time Optimization}}                                                                                                                   & \multicolumn{2}{c}{\textbf{Rolling-horizon Approach}}                                                                                                                             \\ \cline{3-6} 
                  &                              & \multicolumn{1}{c|}{\textbf{\begin{tabular}[c]{@{}c@{}}Computation \\ Time (s)\end{tabular}}} & \textbf{\begin{tabular}[c]{@{}c@{}}Objective\\ Cost (s)\end{tabular}} & \multicolumn{1}{c|}{\textbf{\begin{tabular}[c]{@{}c@{}}Computation \\ Time (s)\end{tabular}}} & \textbf{\begin{tabular}[c]{@{}c@{}}Objective\\ Cost (s)\end{tabular}} \\ \hline
\textbf{Case 1}   & 19                           & \multicolumn{1}{c|}{49.292}                                                                   & 2120.2                                                                & \multicolumn{1}{c|}{31.064}                                                                   & 2120.2                                                                \\ \hline
\textbf{Case 2}   & 25                           & \multicolumn{1}{c|}{169.30}                                                                   & 2298.1                                                                & \multicolumn{1}{c|}{42.414}                                                                   & 2302.2                                                                \\ \hline
\textbf{Case 3}   & 28                           & \multicolumn{1}{c|}{50.514}                                                                   & 2708.0                                                                & \multicolumn{1}{c|}{53.084}                                                                   & 2708.0                                                                \\ \hline
\textbf{Case 4}   & 32                           & \multicolumn{1}{c|}{133.66}                                                                   & 2677.8                                                                & \multicolumn{1}{c|}{67.008}                                                                   & 2677.8                                                                \\ \hline
\textbf{Case 5}   & 36                           & \multicolumn{1}{c|}{784.69}                                                                   & 2736.3                                                                 & \multicolumn{1}{c|}{102.75}                                                                   & 2736.3                                                                \\ \hline
\end{tabular}
}
\end{center}
\end{table}
\subsection{Travel uncertainty impacts under Monte Carlo method}
Even though experiments in Section \ref{sec:exp2} employ rolling horizon and run in a closed-loop with simulator, they do not incorporate disturbances. This absence of disturbances compromises the ability to accurately reflect real-world conditions, thereby limiting the assessment of the algorithm's robustness. Therefore, we add disturbances in the simulator, as described in Section \ref{sec:disturbexp3}. The random values for $\xi_{\text{arr},f}$, $\xi_{\text{hld},f}$ and $\xi_{\text{ob},f}$ are set from 0.05 to 0.2, respectively. We select flight TMA entry times from one busy historical hour, namely, Case 5 in Table \ref{tab:exp2}. For each disturbance scenario, we repeatedly run 100 times under different seed numbers ranging from 0 to 99. 

\begin{figure}[!h]
	\centering
	\includegraphics[width=3.3in]{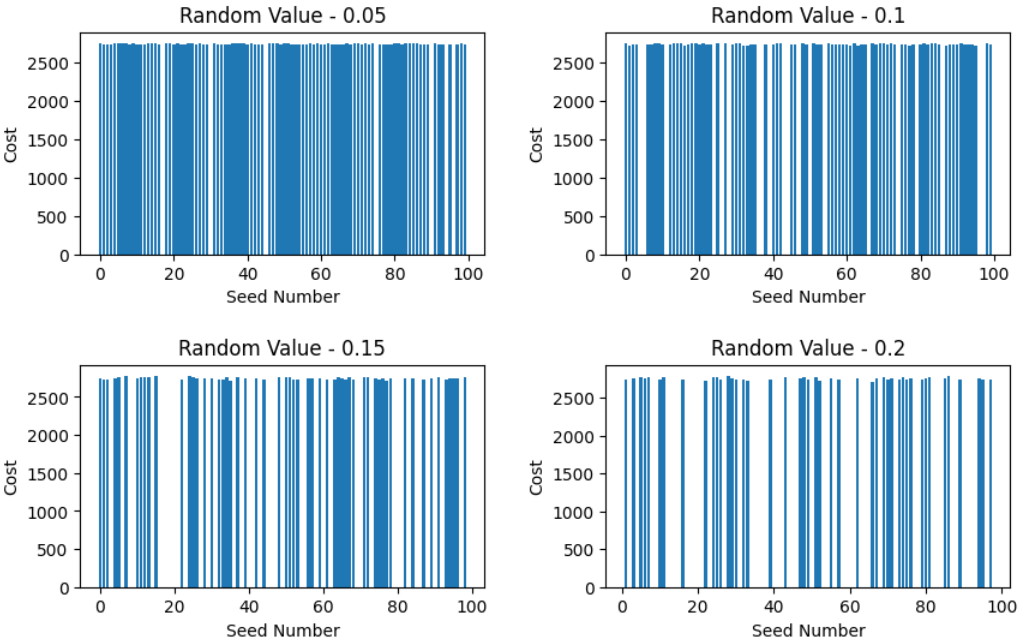}
	\caption{Bar plots of MILP method under 4 random values}
	\label{fig_exp3_barplot}
\end{figure}
\begin{figure}[!ht]
    \centering
    \begin{subfigure}[t]{0.48\columnwidth}
        \centering
        \includegraphics[width=\columnwidth]{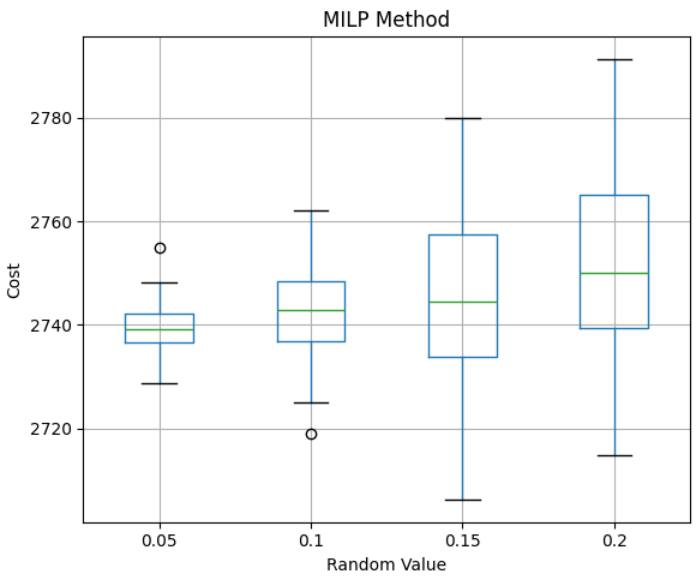}
        \caption{Cost distribution of MILP method under 100 Monte Carlo testings}
        \label{fig_exp3_cost}
    \end{subfigure}%
    ~
    \begin{subfigure}[t]{0.48\columnwidth}
        \centering
        \includegraphics[width=\columnwidth]{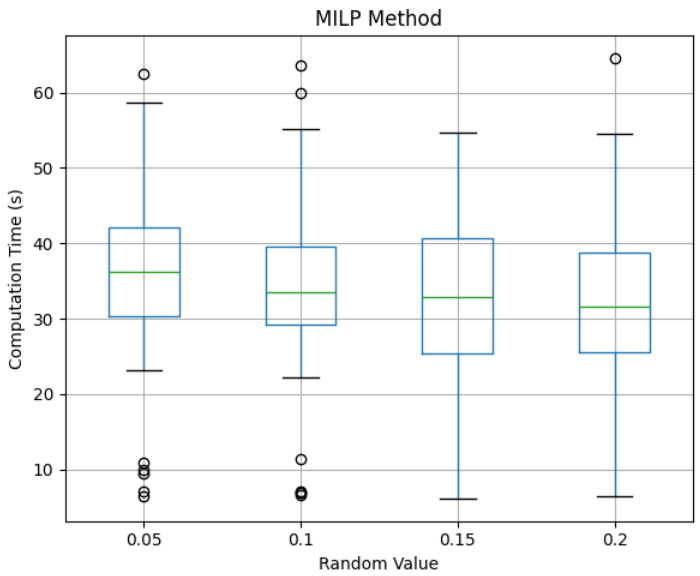}
        \caption{Time distribution of MILP method under 100 Monte Carlo testings}
        \label{fig_exp3_time}
    \end{subfigure}
    \caption{Performance of MILP method}
    \label{fig_exp3_combined}
\end{figure}

Fig. \ref{fig_exp3_barplot} illustrates the cost result (average landing time) under different seed numbers and random values. The empty gap between the blue bars indicates that this run fails to find the feasible solution. Clearly, with the increase of the random value, the gap becomes more and more obvious. Furthermore, to illustrate the cost dynamics in (\ref{eq:cost}), we generated a box plot depicted in Fig. \ref{fig_exp3_cost}, showcasing the mean and variance of costs for all feasible solutions across various random values. We observe a slight increase in the mean cost with increasing random values, accompanied by a substantial rise in variances. This trend suggests that the system becomes more susceptible to fluctuations and uncertainties as the randomness in the environment intensifies.

Moreover, we also provide the computational time of each run in Fig. \ref{fig_exp3_time}. Different from the fluctuations observed in cost, the variation in computational time does not exhibit a clear trend with increasing random values. The majority of runtimes fall within the range of 30 to 40 seconds. This observation suggests that the algorithm's computational efficiency is relatively stable across different random value settings. Furthermore, the reduced computation time observed for Case 5 compared to Table \ref{tab:exp2} is attributed to the adjustment of the control horizon and look-ahead horizon to 600 seconds and 800 seconds, respectively. This adjustment was made to expedite the calculations due to the requirement for 100 runs.

\begin{figure}[!h]
	\centering
	\includegraphics[width=2.5in]{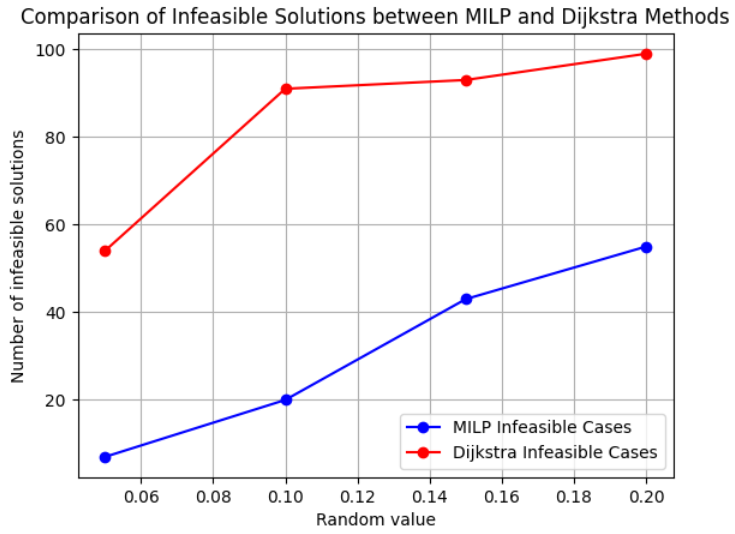}
	\caption{Infeasible solutions under 100 Monte Carlo testings}
	\label{fig_exp3_compare}
\end{figure}

To further illustrate the stability of rolling-based MILP approach, we also conduct the same testings on Dijkstra algorithm. To facilitate integration with the simulator in a closed-loop setup, the rolling horizon approach is also implemented for the Dijkstra method. Fig. \ref{fig_exp3_compare} illustrates the number of infeasible cases for 100 runs under different random values. Clearly, as the random values increase, there is a notable rise in the number of infeasible cases for both algorithms. However, the Dijkstra algorithm demonstrates a considerably higher incidence of infeasible cases compared to MILP, with instances where only a single feasible solution is generated when the random value reaches 0.2. This observation suggests that MILP exhibits greater robustness compared to Dijkstra.
\section{Conclusion} \label{sec:conclusion}
This paper presents a real-time control framework for high-level scheduling in TMA multi-aircraft conflict resolution using a receding horizon strategy. ADS-B data analysis reveals complex TMA behavior, including frequent holdings and vectoring, motivating the prediction of arrival times at the TMA boundary for optimization inputs. The proposed model computes control actions—speed changes, holding times, and route adjustments—to ensure conflict-free operations. Scheduling commands are dynamically updated to resolve conflicts, and an MPC approach with a rolling horizon enables real-time decision-making when integrated with a simulator. The performance of the controller is further evaluated by comparing it to a priority-based Dijkstra algorithm, both in disturbance-free and disturbed environments. The results demonstrate the controller’s superior efficiency, robustness, and ability to handle varying levels of operational complexity, highlighting its potential for practical air traffic management applications.


\section*{Acknowledgment}
This work is supported by the National Research Foundation, Singapore, and the Civil Aviation Authority of Singapore (CAAS), under the Aviation Transformation Programme. (Grant No. ATP2.0 WIC I2R). Any opinions, findings and conclusions, or recommendations expressed in this material are those of the authors and do not reflect the views of the National Research Foundation, Singapore, and the Civil Aviation Authority of Singapore. The authors would like to thank all colleagues from CAAS for providing valuable comments and suggestions on this work.

\bibliographystyle{unsrt}  
\bibliography{ref}  

\end{document}